\documentclass[11pt]{article}
\usepackage{graphicx}
\usepackage{amsmath}
\usepackage{amssymb}
\usepackage{amsxtra}
 \usepackage[english]{babel} 
 \usepackage{latexsym} 
 \usepackage{amsmath} 
 \usepackage{amsfonts} 
 \usepackage{amssymb} 
 \usepackage{textcomp} 
 \usepackage{multirow}
 \usepackage{latexsym}
\usepackage{graphics}
\usepackage{graphicx}
\usepackage{feynmf}
\usepackage{color}
\usepackage{MnSymbol}
\newcommand{\be}{\begin{equation}}
\newcommand{\ee}{\end{equation}}
\newcommand{\ba}{\begin{eqnarray}}
\newcommand{\ea}{\end{eqnarray}}

\newcommand{\pa}{\partial}

\newcommand{\bw}{\begin{widetext}}
\newcommand{\ew}{\end{widetext}}

 \usepackage{graphicx}  
 \usepackage{enumerate} 
 \usepackage{array} 
 \usepackage{tikz}
 \usetikzlibrary{fit,calc,positioning,decorations.pathreplacing,matrix}
 \usepackage[latin3]{inputenc} 
 \usepackage{graphics} 
 \usepackage{color} 
\usepackage{hyperref}

\def\bs{\begin{subequations}}
\def\es{\end{subequations}}

\begin{document} 
\title{Spontaneous symmetry breaking \\
and the Unruh effect \footnote{Paper presented at the Fourth International Conference on the Nature and Ontology of Spacetime, dedicated to the 100th anniversary of the publication of General Relativity, 30 May - 2 June 2016, Golden Sands, Varna, Bulgaria   (Minkowski Institute Press, Montreal 2017).}}
\author{ Antonio Dobado \\
Departamento de  F\'\i sica Te\'orica I\\
 Universidad Complutense, 28040 Madrid, Spain.
}

\maketitle

\begin{abstract}

In this work we consider the ontological status of the Unruh effect. Is it just a formal mathematical result? Or the temperature detected by an accelerating observer can lead to real physical effects such as phase transition. In order to clarify this issue we use the Thermalization Theorem to explore the possibility of having a restoration of the symmetry in a system with spontaneous symmetry breaking of an internal continuous symmetry as seen by an accelerating observer. We conclude that the Unruh effect is an ontic effect, rather than an epistemic one, giving rise, in the particular example considered here, to a phase transition (symmetry restoration)   in the region close to the accelerating observer horizon.
\end{abstract}

\section{Introduction}

Trying to understand better Hawking radiation \cite{Hawking}, Unruh did an amazing discovery in 1976 \cite{Unruh} (see \cite{Crispino:2007eb} for a very complete review). He realized that an observer moving through the Minkowski vacuum 
with a constant acceleration $a$ will detect a thermal bath at temperature:
\be
T=\frac{a \hbar}{2\pi c k_B}.
\ee
This result was first obtained for free bosonic quantum fields  but later it was extended to interacting fields  giving rise to the so called Thermalization Theorem \cite{Lee}. The relevance of the above formula is based, among other things, on the fact that it relates Quantum Mechanics, Relativity and Statistical Physics because it contains the Planck constant $\hbar$, the speed of light $c$ and the Boltzmann constant $k_B$ (in the following we will use natural units with $c= \hbar = k_B= 1$).

There are different approaches to the Unruh effect. The first one is based on 
Bogolyubov transformations, and it was the approach used by the pioners of field quantization on Rindler space \cite {Fulling,Birrel,Parker,Boulware}. There is also the operational approach based in the concept of  Unruh-DeWitt detector where one studies  the response of accelerating detectors to the quantum fluctuations of the fields. Also it is possible to use operator algebra in the context of 
Modular Theory where the concept  of  KMS (Kubo-Martin-Schwinger \cite{Kubo:1957mj}, \cite{Martin:1959jp} )  states plays a dominant role (see   \cite{Earman:2011zz} for detailed review). This is possibly the most abstract approach and the most far away from the physical interpretation of the phenomenon. On the opposite side one can consider the {\it experimental} approach based in analogue systems \cite{Barcelo:2005fc} suggested by Unruh himself 
\cite {Unruh:1980cg},  as for example  by studying the behavior of subsonic-ultrasonic interfaces in Bose-Einstein condensates \cite{Garay:1999sk} . Finally we have the so called Thermalization Theorem. It was introduced by Lee \cite{Lee} and it is  based in a  path integral approach to Quantum Field Theory (QFT) in curved space time. This approach is the most general since it incorporates many elements of the previous ones. In addition it can be applied to any kind of fields like scalars, gauge or fermions and most importantly, to interacting systems.
It also offers a picture of the Unruh effect as an instance of the  Einstein-Podolsky-Rosen Gedanken experiment \cite{Einstein}.
 
 In this work we want to explore the possibility of the Unruh effect for producing non-trivial thermal effects such as phase transitions. For this reason we will be using the Thermalization Theorem approach appropriate for interacting QFT. In particular we will study the symmetry breaking restoration produced by acceleration in the $SO(N+1)$ Linear Sigma Model (LSM). At zero temperature and Minkowsky space this model can (for appropriate values of the parameters) feature  spontaneous symmetry breaking (SSB) from $SO(N+1)$ down to $SO(N)$. The model  is renormalizable and, in addition, it can be solved in a non-perturbative way for large $N$ in a particular limit. Also it has a thermal second order $SO(N+1)$ symmetry restoration at a temperature $T_c=2v \sqrt{3/N}$ in the large $N$ limit, with  $v$ being the vacuum expectation value (VEV). By using the Thermalization Theorem we will show that an accelerating observer will indeed observe a restoration of the symmetry in this model at some critical acceleration:
 \be
 a_c = 4 \pi v \sqrt{\frac{3}{N}}.
 \ee
 Moreover the accelerating observer will see a different value of the VEV at different distances from her horizon so that the restoration of symmetry is produced in the region close to this horizon. Some previous related results have been obtained for the Nambu-Jona-Lasinio  model in   \cite{Ohsaku:2004rv} and  \cite{Ebert:2006bh} 
  and in \cite{Castorina:2012yg}  for the $\lambda \Phi^4$ theory at the one-loop level.

\section{Comoving coordinates for Rindler space}

When dealing with accelerating observers (or detectors) in Minkowski space it is very useful to use Rindler and comoving coordinates. In the four-dimensional Minkowski space $M$ we can introduce
cartesian inertial coordinates $X^\mu =(T,X,Y,Z)$ with metric:
\be
ds^2 = dT^2- dX^2 -dY^2 -dZ^2
\ee
We define also $X_\bot = (Y,Z)$ so that:
\be
ds^2= dT^2-dX^2-dX_\bot^2
\ee
in an obvious notation. Clearly these 
coordinates move along the whole real axis $T,X,Y,Z \in
(-\infty, \infty)$.
Next we can introduce Rindler coordinates as follows:
\ba
T & = & \rho\, \sinh\, \eta   \nonumber \\       
X   &  = & \rho\, \cosh\, \eta        \label{Rindler}
\ea
where $\rho \in (0,\infty)$ and $\eta \in (-\infty,\infty)$. Thus 
these coordinates are covering only the region $X > 
\mid T\mid$, the so called ${\cal R}$ wedge. It is also possible to
introduce the complementary coordinates $\rho'$ and $\eta'$ as:
\ba
T & = &  \rho'\, \sinh\, \eta'    \nonumber \\
X & = & - \rho'\, \cosh\,  \eta'
\ea
covering the left wedge ${\cal L}$ where $- X > \mid T \mid$. In the 
case of the ${\cal R}$  region the metric reads:
\be
ds^2= \rho^2d\eta^2- d\rho^2-dX_\bot^2.
\ee
The other two regions are the origin past ${\cal P}   (T<- \mid X\mid) $ and future  ${\cal F}  (T> \mid X\mid) $.

In Minkowski space, an uniformly accelerating observer in the
$X$ direction will follow a world line like:
\ba   
T(\tau) = \frac{1}{a} \sinh\, (a\tau)    \nonumber \\
X(\tau) = \frac{1}{a} \cosh\, (a\tau) \label{Comoming}
\ea
where we assume $X_\bot$ to be constant and $a>0$ and $\tau$ are the
proper acceleration and the proper time respectively. The same world line is
described in Rindler coordinates by the simple equations:
\ba
\rho  & = & \frac{1}{a}  \nonumber   \\
\eta  & = & a \tau.
\ea
Therefore Rindler coordinates correspond to a network of
observers with different proper constant acceleration $a= 1/
\rho$ and having a clock measuring their proper times in units
of $a\eta$. Those observers have a past and a future horizon at $X=-T$ and
$X=T$ respectively that they find in the infinite remote past  or
future (in proper time)  or also in the limit $\rho
\rightarrow 0$ (infinite acceleration).

In the following it will be very interesting to  introduce a new
system of coordinates on the manifold ${\cal R}$, i.e. $x^\mu=(t,x,y,z)$ defined as:
\ba
T & = & \frac{1}{a}e^{a x}  \sinh\,(a t)  \nonumber  \\
X   &=   &\frac{1}{a}e^{a x}  \cosh\,(a t)  \nonumber  \\
Y  & =  &y    \nonumber  \\
Z  & =  & z.        \label{Comoving}
\ea
These are just the comoving coordinates of a non-rotating
accelerating observer with constant acceleration $a$ in the $X$
direction. Note that $t
,x,y,z \in (-\infty, \infty)$ and one has $\rho= e^{a x}/a$ and
$\eta= a t$. Thus a point with fixed $x$ coordinate is having an
acceleration:
\be
a(x)  = a e^{-a x}
\ee
so that $a(0) = a$ but $a(x)$ goes to infinity when
$x$ goes to $-\infty$ (the horizon) and it goes to zero when $x$
goes to $\infty$. In these  comoving coordinates the metric has
the simple form:
\be
ds^2 = e^{2 a x}(dt^2-dx^2)-dx_\bot^2
\ee
where $x_\bot=(y,z)$.

Alternatively it is possible to define the coordinates
$x'^\mu=(t',x',y',z')$ as:
\ba
T & = & \frac{1}{a}e^{a x'}  \sinh\,(a t')  \nonumber  \\
X   &=   &-\frac{1}{a}e^{a x'}  \cosh\,(a t')  \nonumber  \\
Y  & =  &y '   \nonumber  \\
Z  & =  & z'.
\ea
These coordinates correspond to a comoving observer having a
constant acceleration $a$ along the negative $X$ direction. As
it is easy to show, they can be used to cover the  ${\cal L}$ wedge. In terms of these coordinates the metric has exactly the same
form as in the previous case. (${\cal R}$ wedge). A very important remark
when comparing Rindler coordinates in Eq.\.(\ref{Rindler}) with
comoving coordinates in Eq.\.(\ref{Comoving}) is that Rindler
coordinates do not show any dimensional parameter or physical
scale. In this sense they are similar to Minkowski coordinates.
However comoving coordinates refer to a particular observer with
acceleration $a$ and thus they  depend on this physical scale.
This fact will become relevant later in this work.

\section{The Thermalization Theorem}

The accelerating observer can only feel directly the Minkowski vacuum fluctuations inside ${\cal R}$. However those fluctuations are entangled with the ones corresponding to the left Rindler region  ${\cal L}$ ($    X < -\mid T \mid  $) as in a kind of Einstein,  Podolsky and Rosen setting. The result is that she sees the Minkowski vacuum as a mixed state described by a density matrix $\rho _R$ which, according to the Thermalization Theorem \cite{Lee}, can be written in terms of the Rindler Hamiltonian $\hat H_R $ (the generator of the $t$ time translations) as:  
\be
\hat \rho_R = \frac{ e^{-2\pi \hat H_R/a}    }{Tr e^{-2\pi \hat H_R/a}}.
\ee
Thus the expectation value of any operator $\hat A_R$ defined on the Hilbert space ${\cal H}_R$ corresponding to the region ${\cal R}$ in the Minkowski vacuum $\mid \Omega_M>$ is given by:
\be
< \Omega_M \mid  \hat A_R \mid \Omega_M > = Tr\hat \rho_R \hat A_R.
\ee
This result can be seen as the one found in a  thermal ensemble at temperature $T= a/ 2 \pi$ (in natural units) and it can be understood as a very precise formulation of the Unruh effect. 

In any case one can of course wonder about the ontological status of this effect. Is the above result just formal or does it truly represent a thermal effect? In more prosaic terms: Could it be possible to cook a steak by accelerating it?  More technically speaking: Can the Unruh effect give rise to phase transitions?

\section{The spontaneously broken $SO(N+1)$ Linear Sigma Model}
 
 In order to explore this issue we have considered a model featuring a spontaneous symmetry
  breaking, namely the well known $SO(N+1)$ Linear Sigma Model (LSM). This model is defined in Minkowski space by the Lagrangian:
 \be \label{eq:lsmlagrangian}
\mathcal{L}  = \frac{1}{2} \pa_{\mu} \Phi^T \pa^{\mu} \Phi -V \left(\Phi^T \Phi
 \right)+J \sigma
\ee
where the multiplet $\Phi=(\bar \pi,\sigma)$  contains $N+1$ real 
scalar fields  ($\bar \pi$  is an $N$ component scalar multiplet). The potential is given by:
\be
V \left(\Phi^T \Phi
 \right)=
-\mu^2 \Phi^T \Phi + \lambda \left(\Phi^T \Phi
 \right)^2
\ee
where $\lambda$ is positive in order to have a potential
bounded from below and  $\mu^2$ is positive  
in order to produce  a spontaneous symmetry breaking (SSB). When the external field
is turned off ($J(x)=0$), the SSB pattern is $SO(N+1) \rightarrow SO(N)$ and $N$ Nambu-Goldstone bosons
 appear in the spectrum. 
 
At  the tree level and $a=0$  the low-energy dynamics is controlled by the broken phase where:
\be
 < \Omega_M \mid  \hat{ \pi}^a \mid \Omega_M> =0; \,\, \,\, \,\, < \Omega_M \mid  \hat{ \sigma} \mid \Omega_M> =v.
\ee
and  $v^2=NF^2=\mu^2/2\lambda$. 
Then the relevant degrees 
of freedom are the  $\hat{\pi}$ fields which  
 correspond to the  Nambu-Goldstone bosons (pions). Fluctuations
 along the $\sigma$ direction  correspond to the Higgs,
the massive mode which is relevant at higher energies or temperatures. 

According to the Thermalization Theorem an accelerating observer will see the system
as a canonical ensemble
described by the partition function given by:
\be 
Z_R(a) = Tr e^{-\frac{2\pi}{a} \hat H_R} = \int [d\Phi] \exp
\left( -S_{RE}[\Phi] \right) \ ,
\ee 
with the thermal like  periodic boundary conditions in Euclidean signature:
\be
\Phi(\bar x, 0) = \Phi(\bar x,2\pi/a)
\ee
and also
\be
\Phi(\mid \bar x \mid=\infty , t_E) ^T\Phi(\mid \bar x
\mid=\infty , t_E)= \sigma^2( \mid \bar x \mid=\infty , t_E) =
v^2, 
\ee
where $t_E$ is the Euclidean comoving time.
In comoving coordinates the Euclidean action $S_{RE}[\Phi]$ defined on  ${\cal R}$  is:
\ba
S_{RE}[\bar \pi,\sigma] & = & \frac{1}{2} \int d^4x(     
(\partial_t \bar\pi)^2
+( \partial_t \sigma)^2+   ( \partial_x \bar\pi)^2     
+( \partial_x \sigma)^2    \\
   & + &  \sqrt{g}[ ( \nabla_{\bot} \bar\pi)^2+( \nabla_{\bot} \sigma)^2 + 2 \lambda (\bar \pi^2 +
\sigma^2)^2-2\mu^2 (\bar \pi^2 + \sigma^2) ] )     \nonumber
\ea
 with 
\be
\int d^4
 x= \int_0^{2\pi/a}dt_E\int_{-\infty}^{\infty} dx \int_{-\infty}^{\infty} dy \int_{-\infty}^{\infty} dz 
\ee
and $\sqrt{g}= e^{2ax}$.
In order to compute the partition function in the large $N$
limit a standard technique consists in  introducing an auxiliary scalar field $\phi$ as follows.
The quartic term appearing in the above
partition function is:
\be
 \exp \left( -  \int d^4x \sqrt{g}\lambda(\bar \pi^2 +
\sigma^2)^2 \right).
\ee 
This term can be taken into account just by introducing in the action: 
\be
-\frac{1}{2}\int d^4x \sqrt{g}\left( N \phi^2 - \sqrt{8 \lambda N} \phi
(\bar \pi^2 + \sigma^2)
\right). 
\ee
and performing an additional $[d\phi]$ functional integration after the integrations on the $\bar \pi$ and $\sigma$  fields.

 The (algebraic) Euler-Lagrange equation for $\phi$ simply gives:
 \be
 \phi^2= \frac{2 \lambda}{N }(\bar \pi^2 + \sigma^2)^2
 \ee
 
 Therefore the  partition function can then be written as:
\be 
Z_R(a) = \int [ d\phi]  [d\sigma][d \bar \pi ] \exp \left(
-S_{RE}[\bar \pi, \sigma, \phi]\right).
\ee 
Notice that now all the interactions  are mediated by the
new auxiliary field $\phi$.
In terms of the $\bar \pi$, $\sigma$ and $\phi$
fields the action  reads:
\ba
 S_{RE}[\bar \pi,\sigma,\phi] & = &  \int
d^4 x \sqrt{g}\ [
\frac{1}{2} \pi^a \left( - \square_E - 2\mu^2 + \sqrt{8\lambda
N} \phi \right) \pi^a     \\  
    & +  & \frac{1}{2} \sigma \left( - \square_E - 2\mu^2 +
\sqrt{8\lambda N} \phi \right) \sigma
-  \frac{1}{2 }N \phi^2  ]  . \nonumber
\ea

At this point it is convenient to introduce the new field:
\be
\chi =  4 \lambda (\bar \pi^2 + \sigma^2-v^2)=\phi \sqrt{8 \lambda N}  - 2\mu^2.
\ee
Then the above action becomes:
\ba
 S_{RE}[\bar \pi,\sigma,\chi] & = & \int
d^4 x\sqrt{g} [
\frac{1}{2} \pi^a \left( - \square_E +\chi \right) \pi^a
+ \frac{1}{2} \sigma \left( - \square_E \right)
\sigma    \\
    & + & \frac{1}{2} (\sigma^2- v^2)\chi -\frac{\chi^2}{16
\lambda}-\lambda v^4 ] \nonumber .
 \ea

By performing a standard Gaussian integration of the pion fields we
get:
\be
\int [d\bar \pi ]\exp \left(- \frac{1}{2}\int d^4x \ \sqrt{g} \pi^a
\left[ - \square_E + \chi \right] \pi^a \right)
= \exp \left(  -   \Delta\Gamma[\chi]             \right).
\ee
where:
\be
\Delta\Gamma[\chi] = \frac{N}{2} Tr
\log\frac{-\square_E+\chi}{-\square_E}.
\ee
Thus we have:
\be
Z_R(a)=\int[d\chi] [d\sigma]e^{-\Gamma_R[\sigma,\chi]}
\ee
where the effective action in the exponent is:
\ba
\Gamma_R[\sigma,\chi] & = & \int d^4 x \sqrt{g}[
\frac{1}{2} \sigma \left( - \square_E \right) \sigma
+ \frac{1}{2 }\left( \sigma^2-v^2 \right) \chi   \\    \nonumber
   & -  & \frac{\chi^2}{16\lambda}-\lambda v^4+\frac{N}{2}\log\frac{-\square_E+\chi}{-\square_E} ]
\ea
At the leading order in the large $N$ expansion this is all we
need since we can expand the effective action around some given field configuration 
$\overline{\sigma}$ and $\overline{\chi}$ as:
\be
\Gamma_R[\sigma,\chi]=
\Gamma_R[\overline{\sigma},\overline{\chi}]+\int d^4x
\sqrt{g}\frac{\delta \Gamma_R}{\delta \sigma(x)}\delta
\sigma(x)+\int d^4x \sqrt{g}\frac{\delta \Gamma_R}{\delta
\chi(x)}\delta \chi(x)+ ...
 \ee 
Now we can choose $\overline{\sigma}$ and $\overline{\chi}$ as
the solutions of:
 \ba
\frac{\delta \Gamma_R}{\delta \sigma(x)}   &  = & - \square_E\sigma
+ \chi \sigma =0 \label{Equ1} \\
\frac{\delta \Gamma_R}{\delta \chi(x)} &  =  & \frac{1}{2}\left(
\sigma^2 -v^2 \right)- \frac{\chi}{8
\lambda}+\frac{\delta}{\delta \chi(x)}
\frac{N}{2}   Tr \log\frac{-\square_E+\chi}{-\square_E}=0 \label{Equ2}
\ea
  so that, by using the saddle point approximation:
  \be
Z_R(a)=e^{-\Gamma[\overline{\sigma},\overline{\chi}]}+O(1/N),
\ee
where we have taken into account that $\Gamma[\sigma,\chi]$ is
order $N$. This large $N$ approximation must be understood as $N \rightarrow \infty$ with $\lambda \rightarrow  0$ while keeping $\lambda N$ finite. Then, in this limit  we have:
\ba
\overline{\sigma}(x)   & = &  < \Omega_M \mid\hat \sigma (x) \mid \Omega_M>
 \\ \nonumber
\overline{\sigma}^2(x)   & = & < \Omega_M \mid (\hat \sigma (x) )^2 \mid \Omega_M>.
\ea

\section{The VEV in the Minkoski vacuum
as seen in the comoving frame}

In order to solve the above equations to obtain $\bar \sigma$ and $\bar \chi$ we first realize
that, in the large $\rho$ limit, and keeping $ax<<1$, the
accelerating observer goes into the Minkowski inertial frame which in
turns means that $\overline{\sigma}$ goes to $v$ and
$\overline{\chi}$ goes to zero. In fact those are the boundary
conditions needed for the the applicability of the
thermalization theorem to the system considered here. Therefore  it makes sense trying to solve the
equations in the $\chi=0$ and $ax<<1$ regime. In that case we
have:
  \begin{eqnarray}
0  & = &    \square_E\sigma     \\
0 & = & \sigma^2 -v^2 +\frac{N}{2\pi^3}\int_0^\infty d \Omega 
\frac{\Omega \pi}{2 \rho^2 \tanh(\Omega \pi)}.
   \label{Equ2p2}
\end{eqnarray}

By writing $\Omega$ as $\omega/a$ and using $\rho a = 1+ a
x+...$ we find, up to order $a x$:
\be
\sigma^2=v^2 -\frac{N}{4\pi^2}(1-2ax)\int _0^\infty d \omega
\omega\left( 1 + \frac{2}{e^{\frac{2 \pi}{a}\omega}-1} \right),
\ee
where the first divergent integral requires regularization and renormalization.  This
can be done by using a $x$ dependent ultraviolet
cutoff $\Lambda e^{-ax}$ to compute the divergent integral and performing the renormalization
of the $v$  parameter:
\be
v^2 \rightarrow   v^2-N\frac{\Lambda^2}{2 (2 \pi)^2}(1-2ax+...).
\ee
This renormalization is compatible with the limit $a=0$
(Minkowski inertial coordinates) and with the red/blue shift detected by the accelerating observer
when recieving a signal emmitted at the point $x$. A similar result can be obtained by using dimensional renormalization. In any case 
we have:
\be
\sigma^2=v^2 -\frac{N}{2\pi^2}(1-2ax)\int _0^\infty d
\omega\omega\frac{ 1}{e^{\frac{2 \pi}{a}\omega}-1} +O((ax)^2) .
\ee
By performing the  $\omega $ integration, the Minkowski VEV of the $\hat \sigma^2(x) $ comoving
operator is given in the $ax<<1$ regime by:
\be
\bar
\sigma^2(x)=< \Omega_M \mid (\hat \sigma (x) )^2 \mid \Omega_M>=v^2\left( 1- \frac{a^2N}{12 (2\pi)^2 v^2}(1-2ax)
\right).
 \ee  
By introducing the critical acceleration:
\be
a_c^2= 3(4 \pi)^2 \frac{v^2}{N}
\ee
we have:
 \be
\bar \sigma^2(x)=v^2\left( 1- \frac{a^2}{a_c^2}+
2\frac{a^3}{a_c^2}x+... \right) \label{linear}.
 \ee 
Notice that at this order this is also a solution of
Eq.~(\ref{Equ1}). Therefore, at the origin of the accelerating    
frame ($x=0$ or $\rho=1/a$), the squared VEV of the $\hat \sigma$
field is given by:
\be
\bar \sigma^2(0)=< \Omega_M \mid (\hat \sigma (0) )^2 \mid \Omega_M>=v^2\left(
1- \frac{a^2}{a_c^2} \right)
 \ee  
 for $0 \le a \le a_c$ and clearly:  
   \be
< \Omega_M \mid   (\hat \sigma (0) )^2 \mid \Omega_M>=0
 \ee  
for $a > a_c$. This is exactly the thermal behavior of the LSM
in the large $N$ limit with $a/a_c$ playing the role of $T/T_c$
(as seen by a  inertial observer).
It corresponds to a second order phase transition at the critical
acceleration $a_c$ where the original spontaneously broken
symmetry is restored for the accelerating observer.

Now let us consider a different accelerating observer at Rindler
coordinate $\rho' = 1/a'$. This observer will find a similar
result just changing $a$ by $a'$. From the point of view of the
first observer the second observer is located at some point $x'$
given by:
\be
\rho'=\frac{1}{a'}=\frac{1}{a}e^{a x'}
\ee
i.e. the acceleration of the second observer is $a'=a e^{-ax'}$.
In this way it is immediate to find the position dependent
result for the squared VEV of the $\sigma$ field which, in comoving coordinates, is given by:
\be
\bar \sigma^2(x)=< \Omega_M \mid   (\hat \sigma (x) )^2 \mid \Omega_M>=v^2\left( 1- \frac{a^2}{a_c^2}e^{-2ax} \right)
\label{result}
\ee
or, in Rindler coordinates, by:
\be
\bar \sigma^2(\rho)=v^2\left( 1- \frac{1}{ a_c^2\rho^2}
\right).
\ee

 \section{The VEV landscape}
       
Therefore, according to Eq.~(\ref{result}), the $\sigma$ field VEV
seen by the accelerating (comoving) observer is position
dependent. This is not strange since the proper acceleration
along the $x$ direction is breaking the Minkowski translation
(and rotation) invariance. Now let us assume a comoving frame
acceleration $a$ belonging to the interval $0 < a < a_c$. 
The squared VEV is a function on the coordinate $x$ ranging
from $v^2$ for $x=\infty$ to zero, which is reached at some
negative $x$ value given by:
\be
x_c=- \frac{1}{2a}\log \frac{a_c^2}{a^2  } < 0
\ee
where the phase transition takes place. Notice that the locus
$x=x_c$ is indeed a surface because of the two other spatial
dimensions $x_\bot$ which are free since the VEV is $x_\bot$ (as
well as $t$) independent.

By using the approximation in  Eq.~(\ref{linear}) one finds:
\be
x_c\simeq-\frac{1}{2a}(\frac{a_c^2}{a^2}-1)<0.
\ee
In this case one has to consider also a second critical
value $x=x_c'$ where the squared VEV equals the asymptotic value
$v^2$:
\be
x_c'= \frac{1}{2a}>0.
\ee
Obviously this approximation is useful only in the region
$x_c<x<x_c'$ at most.

\begin{figure}[tb]
  \begin{center}
    \includegraphics[width=0.4\textwidth]{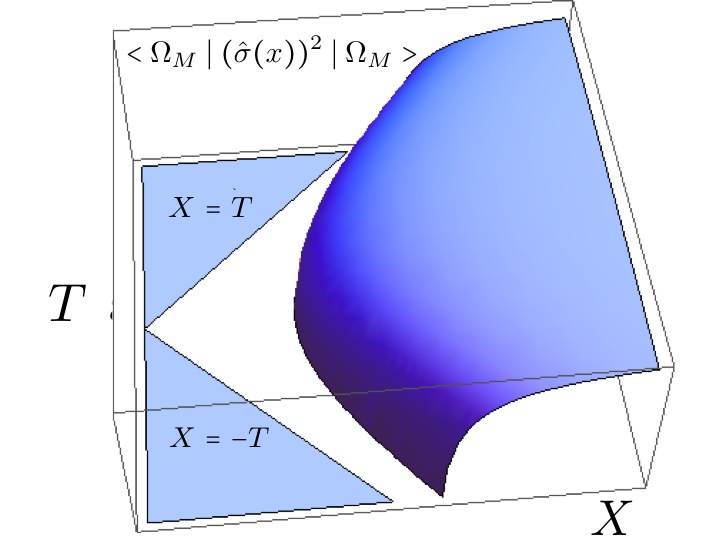}
    \caption{\label{fig:landscape} VEV of  $\hat \sigma^2$  in Minkowski vacuum for different points of the space-time as seen by the accelerating observer}
  \end{center}
\end{figure}

Now it is possible to write to $\overline \sigma^2$ in terms
of the Minkowski coordinates $X$ and $T$:
\be
\bar \sigma^2=v^2\left( 1- \frac{1}{ a_c^2\rho^2}   \label{VEV1}
\right)=v^2\left( 1- \frac{1}{ a_c^2(X^2-T^2)} \right).
\ee
It is very interesting to realize that this function does not
depend on the acceleration $a$ but only on $v$ and the critical
acceleration $a_c$ (which depends only on $v$ and on $N$). In
other words the VEV landscape depends only on the parameters
defining the LSM, but not on the acceleration of the comoving
observer. In Fig.~\ref{fig:landscape} we can find a plot of the VEV on the Minkowski space as seen by the accelerating observer.

On the other hand we also have for the Minkowski quantum field $\hat \sigma_M(X)$:
\be
< \Omega_M \mid   (\hat \sigma_M (X) )^2 \mid \Omega_M > =v^2. \label{VEV2}
\ee
At this point one may wonder; as $\sigma$ is an scalar and, at the classical level, one should have:
\be
\sigma(x)= \sigma_M(X)
\ee
on ${\cal R}$. Is this not in contradiction with Eq.\.(\ref{VEV1}) and Eq.\.(\ref{VEV2})? The answer clearly is not, since:
\be
< \Omega_M \mid   (\hat \sigma_M (X) )^2 \mid \Omega_M> \ne < \Omega_M \mid   (\hat \sigma (x) )^2 \mid \Omega_M>. \label{ineq}
\ee
The reason is that  $\hat \sigma_M (X)$ is an operator defined on the Minkowski Hilbert space ${\cal H}_M={\cal H}_L  \otimes {\cal H}_R$ where
${\cal H}_L $ and  $ {\cal H}_R $ are the Hilbert spaces corresponding to the regions ${\cal L}$  and ${\cal R}$ respectively. However 
 $\hat \sigma (x)=\hat \sigma_R (x)  $ is an operator defined only on  ${\cal  H}_R$, and it must be understood as $1 \otimes \hat \sigma_R (x) $ when acting on $ \mid \Omega_M >$. An event belonging to the region ${\cal P}$ can affect events both in ${\cal L}$ and ${\cal R}$. Thus if $X_L \in{\cal  L}$ and $X_R \in{\cal  R}$, $< \Omega_M  \mid \hat \sigma_M (X_L) \hat \sigma_M(X_R) \mid \Omega_M> $ does not necessarily  vanish. This shows that that $\sigma_M$ is not the tensorial product of 
$\sigma_L$ and $\sigma_R$  i.e.
\be
\hat \sigma_M(X) \ne \theta(-X)\hat \sigma_L(x)  \otimes  \theta(X)\hat \sigma_R(x)
\ee
and  then Eq.\.(\ref{VEV1}) and Eq.\.(\ref{VEV2}) are not incompatible at all.
\section{Conclusions}

The Unruh effect is an unavoidable consequence of QFT for accelerating observer. It applies to interacting theories and to any kind of fields (scalar, fermionic, gauge, etc). It can give rise to collective non-trivial phenomena such as phase transitions. In particular, in this work we have shown  that a continuous spontaneously broken symmetry  is restored for an accelerating observer. For her the VEV of the field depends on the position and it vanishes beyond a surface in the horizon direction. We conclude that all these facts are a solid evidence in favor of the ontic character of the Unruh effect.

\section*{Acknowledgments}
The author thanks Vesselin Petkov and all the participants at the Fourth International Conference on the Nature and Ontology of Spacetime, Varna, Bulgaria (2016). Work supported by Spanish grants MINECO:FPA2014-53375-C2-1-P  and FPA2016-75654-C2-1-P.     

\newpage

 

\end{document}